\documentclass[aps,reprint]{revtex4-2}
\usepackage[utf8]{inputenc}
\usepackage[english]{babel}
\usepackage{amsmath}
\usepackage{amssymb}
\usepackage{float}
\usepackage{graphicx}
\usepackage{hyperref}

\begin{document}

\title{\textbf{Electron-beam-induced adatom-vacancy-complexes in mono- and bilayer phosphorene}}
\author{Carsten Speckmann$^{1,2,*}$, Andrea Angeletti$^{1,2}$, Lukáš Kývala$^{1,2}$,\\ David Lamprecht$^{1,3}$, Felix Herterich$^1$, Clemens Mangler$^1$, Lado Filipovic$^3$, Christoph Dellago$^1$, Cesare Franchini$^1$, and Jani Kotakoski$^{1,*}$\\
$^1$Faculty of Physics, University of Vienna, Boltzmanngasse 5,\\ 1090 Vienna, Austria\\
$^2$Vienna Doctoral School in Physics, University of Vienna, Boltzmanngasse 5,\\ 1090 Vienna, Austria\\
$^3$Institute for Microelectronics, TU Wien, Gußhausstraße 27-29/E360, \\ 1040 Vienna, Austria\\
$^*$Email: carsten.speckmann@univie.ac.at and jani.kotakoski@univie.ac.at}
\date{\today}


\begin{abstract}

    Phosphorene, a puckered two-dimensional allotrope of phosphorus, has sparked considerable interest in recent years due to its potential especially for optoelectronic applications with its layer-number-dependant direct band gap and strongly bound excitons. 
    However, detailed experimental characterization of its intrinsic defects as well as its defect creation characteristics under electron irradiation are scarce.
    Here, we report on the creation and stability of a variety of defect configurations under $60$~kV electron irradiation in mono- and bilayer phosphorene including the first experimental reports of stable adatom-vacancy-complexes. 
    Displacement cross section measurements in bilayer phosphorene yield a value of $7.7 \pm 1.4$~barn with an estimated lifetime of adatom-vacancy-complexes of $19.9 \pm 0.7$~s, while some are stable for up to $68$~s under continuous electron irradiation.
    Surprisingly, {\it ab initio}-based simulations indicate that the complexes should readily recombine, even in structures strained by up to $3$~\%.
    The presented results will help to improve the understanding of the wide variety of defects in phosphorene, their creation, and their stability, which may enable new pathways for defect engineered phosphorene devices.

\end{abstract} 

\maketitle
\newpage

\section*{Introduction}
	
	Semiconducting two-dimensional (2D) materials have been a research focus for some time as they provide the potential to create smaller, and more importantly, the thinnest possible devices in nanoelectronics~\cite{lin_emerging_2020, cheng_improving_2022}.
	One of the most interesting materials is phosphorene, a single layer of black phosphorus (BP), a layered material where layers show a puckered, hexagonal structure with lattice parameters of $3.3$~\r{A} and $4.53$~\r{A} along the zig-zag (ZZ) and armchair (AC) directions, respectively~\cite{kou_phosphorene_2015}.
	BP is also the thermodynamically most stable allotrope of phosphorus~\cite{housecroft_inorganic_2004}.
	Mono- or few-layer phosphorene has recently received increasing attention for its electronic and optoelectronic application potential~\cite{xia_rediscovering_2014, ling_renaissance_2015, kou_phosphorene_2015, carvalho_phosphorene_2016, wan_recent_2020}.
    This is due to its thickness-dependent direct band gap, which ranges from $1.73$~eV in the monolayer to $0.35$~eV in the bulk~\cite{li_direct_2017}, mechanical flexibility~\cite{wei_superior_2014}, and its strongly bound excitons with binding energies as high as $0.9$~eV~\cite{wang_highly_2015}.  
	More importantly, due to its unique atomic structure, phosphorene shows interesting anisotropic properties, such as Raman mode sensitivity on polarization angle and sample orientation~\cite{wu_identifying_2015}, an anisotropic charge carrier effective mass~\cite{qiao_high-mobility_2014, low_tunable_2014}, a negative Poisson's ratio \cite{jiang_negative_2014}, anisotropic bright excitons~\cite{tran_layer-controlled_2014, wang_highly_2015}, as well as an anisotropic electrical and thermal conductance~\cite{fei_enhanced_2014, qin_thermal_2018} that could all be used for tailored applications.
	However, phosphorene is extremely sensitive to oxidation, leading to rapid degradation within hours or even minutes of being exposed to ambient conditions~\cite{pei_producing_2016, kuriakose_black_2018}, complicating both the interpretation of measurements where samples were exposed to oxygen during the process as well as its applications potential.
	
	It is well known that defects influence the properties of materials, which is why a thorough understanding of their creation and effects of these defects is needed.
	Density-functional theory (DFT) calculations for the knock-on displacement thresholds in phosphorene under electron-irradiation indicate that beam damage would only occur with an elastically transferred kinetic energy of $\geq 6$~eV for a pristine sample and $\geq 4.5$~eV for a sample with defects~\cite{vierimaa_phosphorene_2016}; however, experimentally, this has not been studied yet.
	In addition, recent studies show that in semiconducting and insulating 2D materials, beam damage can occur below the knock-on threshold due to electronic excitations~\cite{kretschmer_formation_2020, speckmann_combined_2023, bui_creation_2023}.
	This can also be expected to take place with phosphorene.
	In the case of vacancies, theory predicts monovacancies to induce a magnetic moment of $1$~$\mu_B$~\cite{srivastava_tuning_2015, hu_geometric_2015} as well as an in-gap state~\cite{liu_two-dimensional_2014, srivastava_tuning_2015, gaberle_structure_2018}, whereas divacancies are predicted to show neither~\cite{liu_two-dimensional_2014, srivastava_tuning_2015}.
	Defects have also been suggested to serve as nucleation sites for surface oxidation when exposed to oxygen or water vapor~\cite{kuntz_control_2017}, while, in combination with strain, they can also lead to a structural phase transitions serving as catalysing sites for new phases~\cite{chen_strain_2018}.
	Experimental characterization of defects in mono- and few-layer phosphorene so far was mainly done using scanning tunnelling microscopy (STM)~\cite{zhang_surface_2009, kiraly_probing_2017, riffle_impact_2018, fang_electronic_2022, qiu_resolving_2017, wentink_substitutional_2021, harsh_identification_2022} and transmission electron microscopy (TEM)~\cite{lee_atomic-scale_2017, yao_situ_2020, lee_fabrication_2020, lee_atomically_2022}.
	A layer-by-layer thinning of multilayer phosphorene under $80$~kV electron irradiation was reported for pristine samples~\cite{lee_atomic-scale_2017, lee_atomically_2022} and graphene/phosphorene heterostructures~\cite{lee_fabrication_2020} where for all cases, a (reconstructed) ZZ edge configuration during the thinning process was found to be preferred over an AC configuration.
	Furthermore, a detailed study of specific defect configurations in mono- and few-layer phosphorene at elevated temperatures of $523–673$~K has reported the presence of divacancies, tetravacancies, and long defect lines, along with their most stable configurations, diffusion behavior, as well as possible transformation mechanisms~\cite{yao_situ_2020}.
	Experimental observations of monovacacies in TEM are limited to a recent study claiming to have observed a monovacancy for the first time with this method~\cite{rabiei_baboukani_defects_2022}, whereas several studies were already published on the observations of monovacancies using STM~\cite{zhang_surface_2009, kiraly_probing_2017, riffle_impact_2018, fang_electronic_2022}.
	However, similar features in STM have also been attributed to substitutional tin impurities rather then phosphorus monovacancies~\cite{qiu_resolving_2017, wentink_substitutional_2021, harsh_identification_2022}.
			
    In this study, we use scanning transmission electron microscopy (STEM) high angle annular dark field (HAADF) imaging to record the effects of a $60$~keV electron beam on a bilayer phosphorene sample, characterizing the created defects and their dynamics.
    In addition to previously reported vacancy configurations, we further identify a number of yet unreported defects.
    Notably, almost all vacancies show adjacent phosphorus interstitials or adatoms, which is the first observation of such defects in phosphorene.
    DFT was used to relax models of the observed structures followed by STEM simulations to verifying their correct identification.
    We estimate the displacement cross section of bilayer phosphorene under $60$~keV electron irradiation to be $7.7 \pm 1.4$~barn with an estimated lifetime of adatom-vacancy-complexes under electron irradiation of $19.9 \pm 0.7$~s and a maximum observed lifetime of up to $68$~s.
    Additionally, we give a rough estimate for the cross sections of the individual phosphorene sublayers towards and away from the electron source for the farther phosphorene layer of the bilayer to be $2.0 \pm 0.3$~barn and $10.4 \pm 1.6$~barn, respectively. 
    For the other layer, much fewer defects were observed.
    Molecular dynamics simulations, accelerated by high-dimensional neural network potentials (HDNNP-MD), were employed to asses the stability of the adatom-vacancy-complexes both under equilibrium conditions, as well as under isotropic and anisotropic strain/stress of up to 3~\%.
    However, all of these simulations resulted in the recombination of adatoms and defects far below the observed lifetimes.

\section*{Results and Discussion}
	
\begin{figure*}[ht!]
\centering
\includegraphics[width=0.6\textwidth]{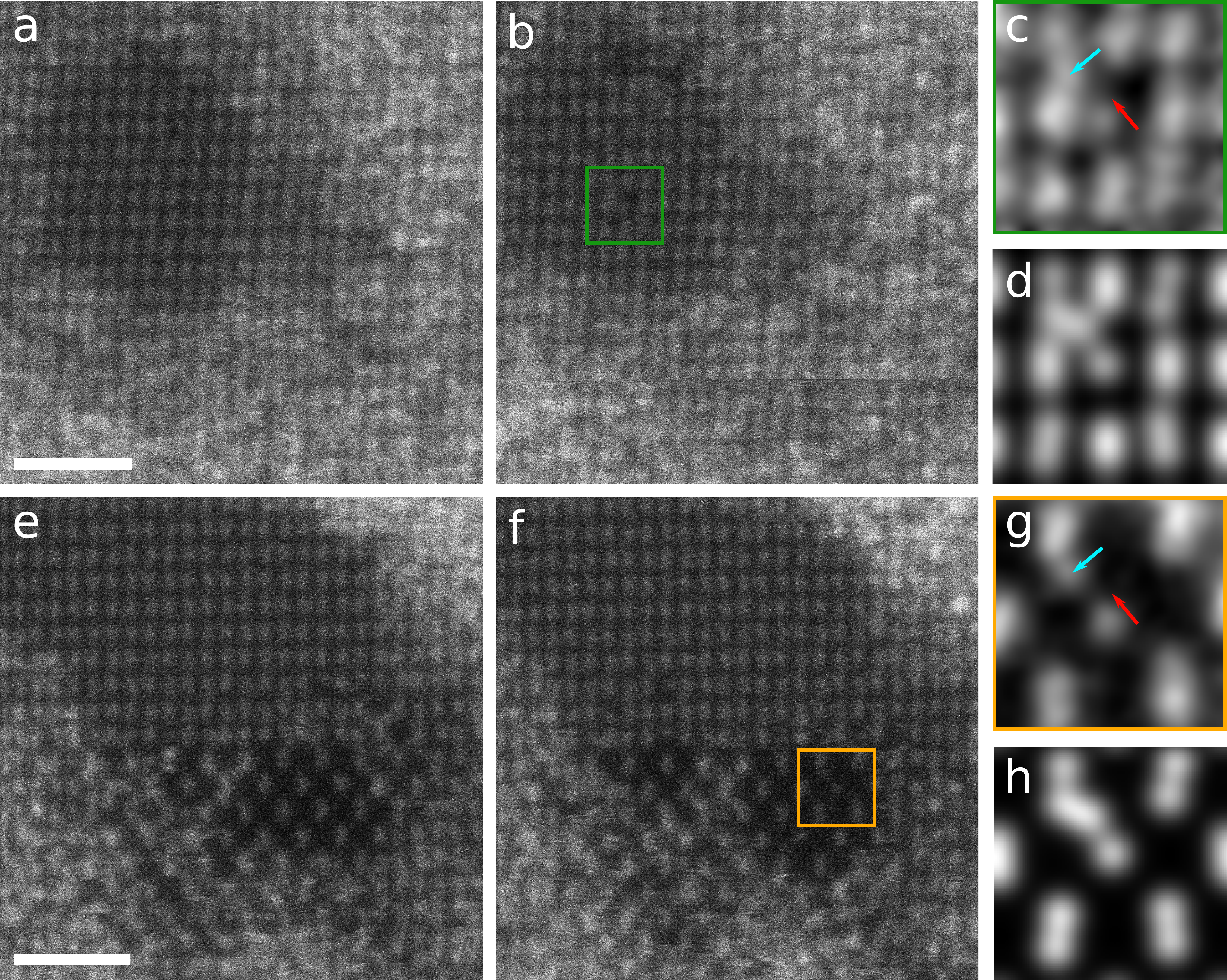}
    \caption{{\bf Creation of adatom-vacancy-complexes in bi- and monolayer phosphorene.} 
    Two STEM-HAADF image pairs of bilayer phosphorene (a-b) and a monolayer patch of phosphorene (e-f) recorded with $1024~\mathrm{px}\times 1024~\mathrm{px}$ and a frame time of $\sim$ $8.5$~s. 
    A zoomed-in image of the area marked with a green (orange) square in b (f) after applying a Gaussian blur of $6$~px ($4$~px) is shown in c (g) where cyan and red arrows denote adatoms and vacancies, respectively.
    Simulated STEM-HAADF images of an adatom-vacancy-complex in bilayer (monolayer) phosphorene are shown in d (h).
    The scale bars are $1$~nm.}
\label{fig:adatoms}
\end{figure*}

	The samples were mechanically exfoliated in an Ar atmosphere and brought into vacuum without being exposed to air.
	For details on the sample preparation, we refer to the Methods section.
	A typical clean area of the sample can be seen in Fig.~\ref{fig:adatoms}a, showing an AB stacked bilayer of phosphorene surrounded by (brighter) contamination.
	The exact composition of this type of contamination is unknown, but previous studies indicate a mixture of different hydrocarbons~\cite{li_effect_2013, inani_silicon_2019, palinkas_composition_2022, tilmann_identification_2023} as well as polymer residue from the transfer process~\cite{schwartz_chemical_2019, tilmann_identification_2023}.
	The effect of electron irradiation on this structure can be seen in Fig.~\ref{fig:adatoms}b, where defects were created in the area marked with the green square after $\sim$ $8.5$~s.
	The area within the green square is magnified in Fig.~\ref{fig:adatoms}c where a red arrow highlights a missing phosphorus atom and a cyan arrow a phosphorus interstitial or adatom.
	To verify the interpretation of this feature as an additional phosphorus atom which is located next to a vacancy, which from now on will be called an adatom-vacancy-complex, we used DFT to first relax a model of the observed structure followed by STEM image simulations. 
    We find a good agreement between the simulated STEM image of the relaxed structure (Fig.~\ref{fig:adatoms}d) and the experimental image.
    Details on the relaxation and the simulation process can be found in the Methods section.
	To our knowledge, this is the first experimental verification of an adatom-vacancy-complex in phosphorene, which has previously only been predicted by theory~\cite{vierimaa_phosphorene_2016}.
	Previous (S)TEM studies on defects in phosphorene~\cite{lee_atomic-scale_2017, yao_situ_2020, lee_fabrication_2020, lee_atomically_2022} never mentioned the existence of phosphorus adatoms, which might be due to the lower electron acceleration voltage of $60$~kV used here and/or due to the fact that the sample was not heated during the experiments in this study and/or due to the low pressure in the objective area of the microscope ($\sim$~10$^{-10}$~mbar).
	
	In Fig.~\ref{fig:adatoms}e-h, starting with panel e, we show a similar behaviour for a small patch of monolayer phosphorene on the bottom of the image. 
	This monolayer flake was created by prolonged electron irradiation, utilizing the previously reported layer-by-layer thinning~\cite{lee_atomic-scale_2017, lee_atomically_2022}.
    In most cases, this process created a pore before a monolayer became visible, but occasionally resulted in a monolayer, allowing its imaging.

	In Fig.~\ref{fig:adatoms}g, showing a magnified view of the area marked with an orange square in Fig.~\ref{fig:adatoms}f, an adatom-vacancy-complex that was created via electron irradiation can be seen.
	Those were again verified using DFT and STEM image simulations, visible in Fig.~\ref{fig:adatoms}h, where we can see an excellent agreement between experiment and simulation.
	Given the low theoretical energy barrier for point defect diffusion in phosphorene (monovacancies: $0.09$~eV~\cite{vierimaa_phosphorene_2016} - $0.25$~eV~\cite{gaberle_structure_2018} along ZZ direction; $0.38$~eV~\cite{hu_geometric_2015} - $0.44$~eV~\cite{gaberle_structure_2018} along AC direction; adatoms: $0.15$~eV~\cite{gaberle_structure_2018} - $0.23$~eV~\cite{vierimaa_phosphorene_2016} along ZZ direction; $\gtrsim$~$1.19$~eV~\cite{vierimaa_phosphorene_2016} along AC direction) and therefore their expected extreme mobility at room temperature, it is surprising to find a phosphorus monovacancy and phosphorus adatom to be stable for the several milliseconds, required to record its feature in this image.
	This is especially intriguing, as there is only one experimental TEM study claiming to have observed a monovacancy in phosphorene~\cite{rabiei_baboukani_defects_2022}, with unfortunately somewhat ambiguous images.

\begin{figure*}[ht!]
\centering
\includegraphics[width=0.6\textwidth]{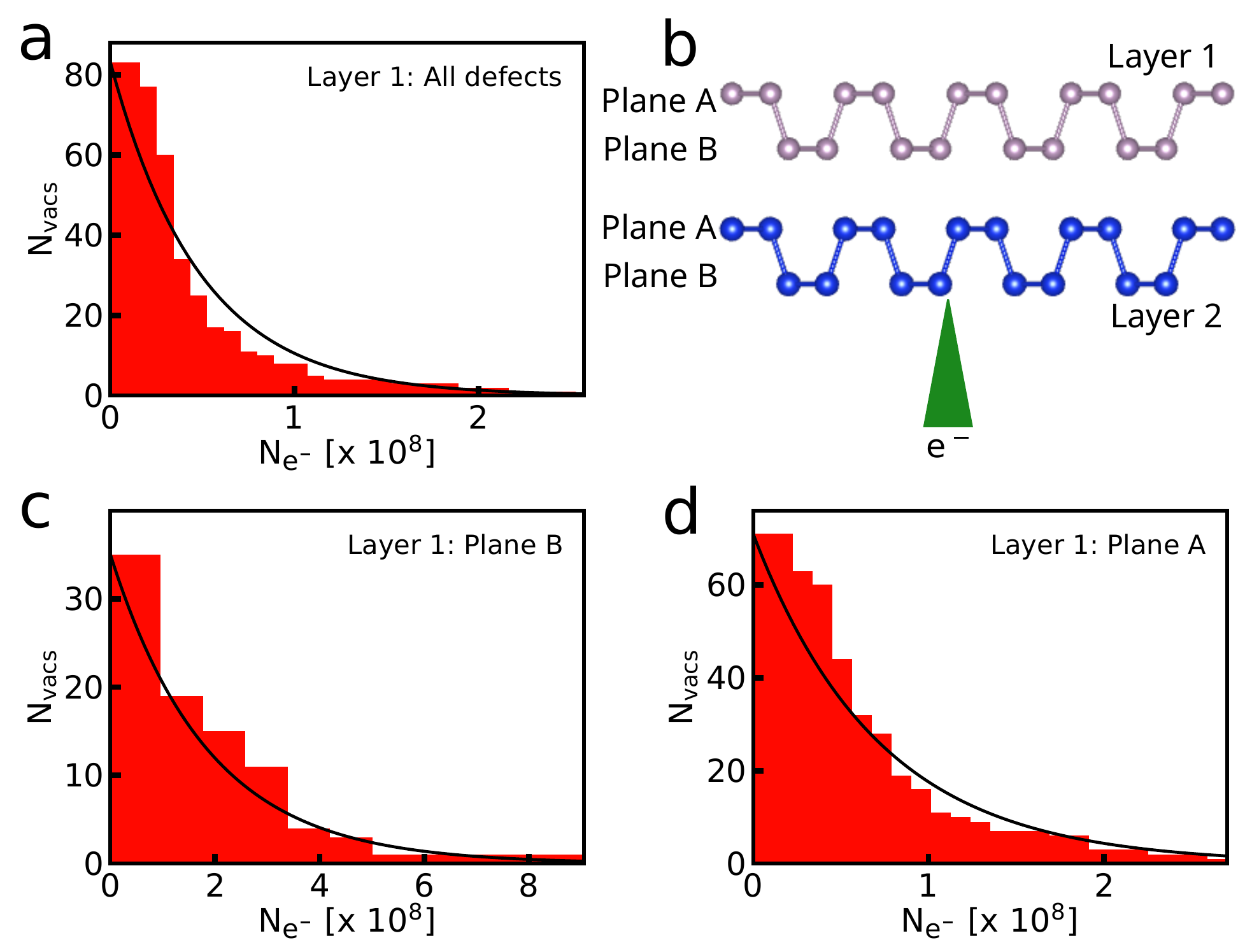}
    \caption{{\bf Statistics of the number of created defects in bilayer phosphorene.} 
    Panel b shows a schematic side view illustration of the phosphorene bilayer structure, indicating the different layers (colored in light purple and blue) and planes in the structure with respect to the incident electron beam. 
    Panels a,c,d exhibit the cumulative statistics for the number of electrons needed for the creation of a single defect including all defects of layer 1 in panel a, only the defects in one plane (B) of layer 1 in panel c and those of the other plane (A) in panel d.
    Here, each bin indicates the number of defects created with an electron dose higher than the current one. 
    Additionally, an exponential decline (shown in black) was fitted to the data (see text).}
\label{fig:stats}
\end{figure*}

	To further estimate the probability of creating a phosphorus vacancy in the phosphorene bilayer, we identified all defects that were created during our measurements and correlated those with the electron doses that were applied to the sample between each appearance of a new defect.
	We then used Poissonian statistics as described in Ref.~\cite{susi_silicon--carbon_2014} to get a mean number of electrons needed for the creation of a single defect (see Fig.~\ref{fig:stats}a).
	In the shown bar plots, at each electron dose the number of defects that required a higher electron dose to be created is shown.
	According to Poisson statistics, this should show an exponential decline, as is the case here.
	The mean electron dose corresponding to the data for all defects in layer 1 yields a cross section of $7.7 \pm 1.4$~barn.
	
	Note that most defects observed during these experiments were found in one of the two layers, indicating that there are different cross sections for the individual layers.
	Following the previous argument by Vierimaa et al.~\cite{vierimaa_phosphorene_2016}, it is justified to assume that the layer hosting most of the defects is the one further away from the electron source (labeled layer 1 in Fig.~\ref{fig:stats}b).	
	To estimate the cross sections of the individual phosphorene sublayers (labeled planes A and B in Fig.~\ref{fig:stats}b), we assigned the defects observed in the experiments to their distinct planes to test the hypothesis that one of the planes has a significantly higher cross section compared to the other.
	The results and statistical analysis are shown in Fig.~\ref{fig:stats}c,d.
	Possible deviations from perfect Poissonian statistics could be explained by the migration of some defects between the planes thermally or due to the electron beam.
	The mean values obtained via these fits yield cross sections for the two planes of $2.0 \pm 0.3$~barn and $10.4 \pm 1.6$~barn for panel c and d, respectively.

    \begin{figure*}[ht!]
    \centering
    \includegraphics[width=0.6\textwidth]{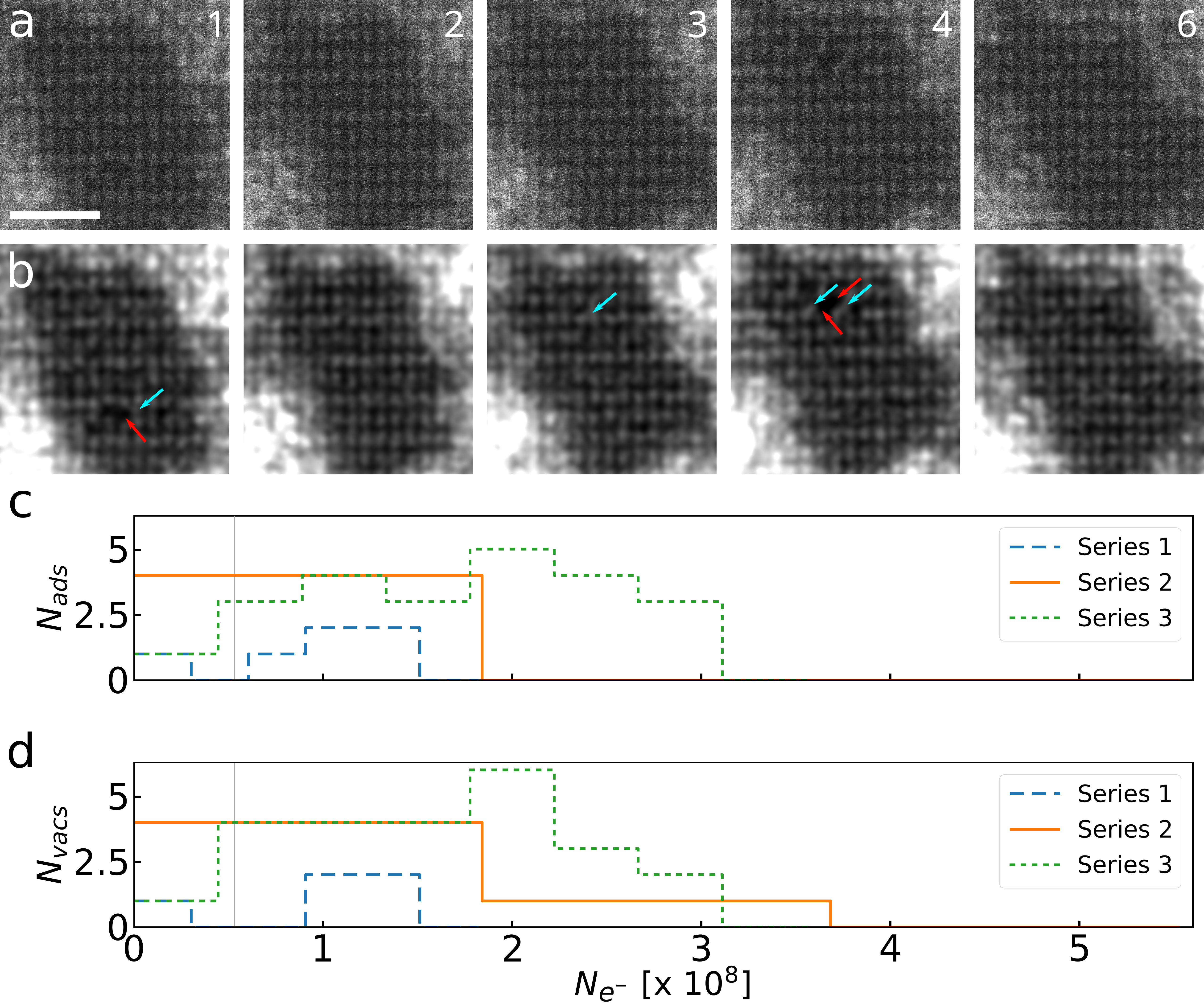}
    \caption{{\bf STEM-HAADF image series of bilayer phosphorene.} 
    Panel a shows five as-recorded images of an image series recorded with $512~\mathrm{px}\times 512~\mathrm{px}$ and a frame time of $\sim$ 2.2~s, where the frame number is indicated in the top right corner of each image. 
    Frame 5 was omitted as it displayed the same structure as frame 4.
    A Gaussian blur ($4$~px) was applied to the images in panel b, and cyan and red arrows were added to indicate adatoms and vacancies, respectively.
    Panels c and d show the evolution of the number of adatoms and vacancies as a function of the total number of electrons impinging on the sample for three example image series.
    The vertical grey lines indicates the average number of electrons needed to create one defect in bilayer phosphorene.
    The dashed blue line labeled "Series 1" corresponds to the images shown in a and b.
    The scale bar is $1$~nm.}
    \label{fig:healing}
    \end{figure*}
	
	It is possible to estimate the displacement energy threshold based on these cross sections by assuming a pure knock-on process using the McKinley-Feshbach-formalism~\cite{mckinley_coulomb_1948}, while also considering the lattice vibrations~\cite{meyer_accurate_2012, meyer_erratum_2013, susi_isotope_2016, chirita_three-dimensional_2022} using the Debye model with a Debye temperature of $278.66$~K for phosphorene~\cite{qin_anisotropic_2015}.
	Those result in a threshold value of $4.79 \pm 0.03$~eV for the whole structure while the plane cross sections yield $5.01 \pm 0.03$~eV and $4.73 \pm 0.03$~eV  for plane B and A, respectively.
	Those values are considerably lower than previously reported theoretical displacement threshold energies of $10.5$~eV in pristine phosphorene or $7$~eV in phosphorene with adatom-vacancy-complexes~\cite{vierimaa_phosphorene_2016}, which suggests that inelastic scattering effects can not be neglected for phosphorene, similar to MoS$_2$ and hBN~\cite{kretschmer_formation_2020, speckmann_combined_2023, bui_creation_2023}.

    \begin{figure*}[ht!]
    \centering
    \includegraphics[width=0.6\textwidth]{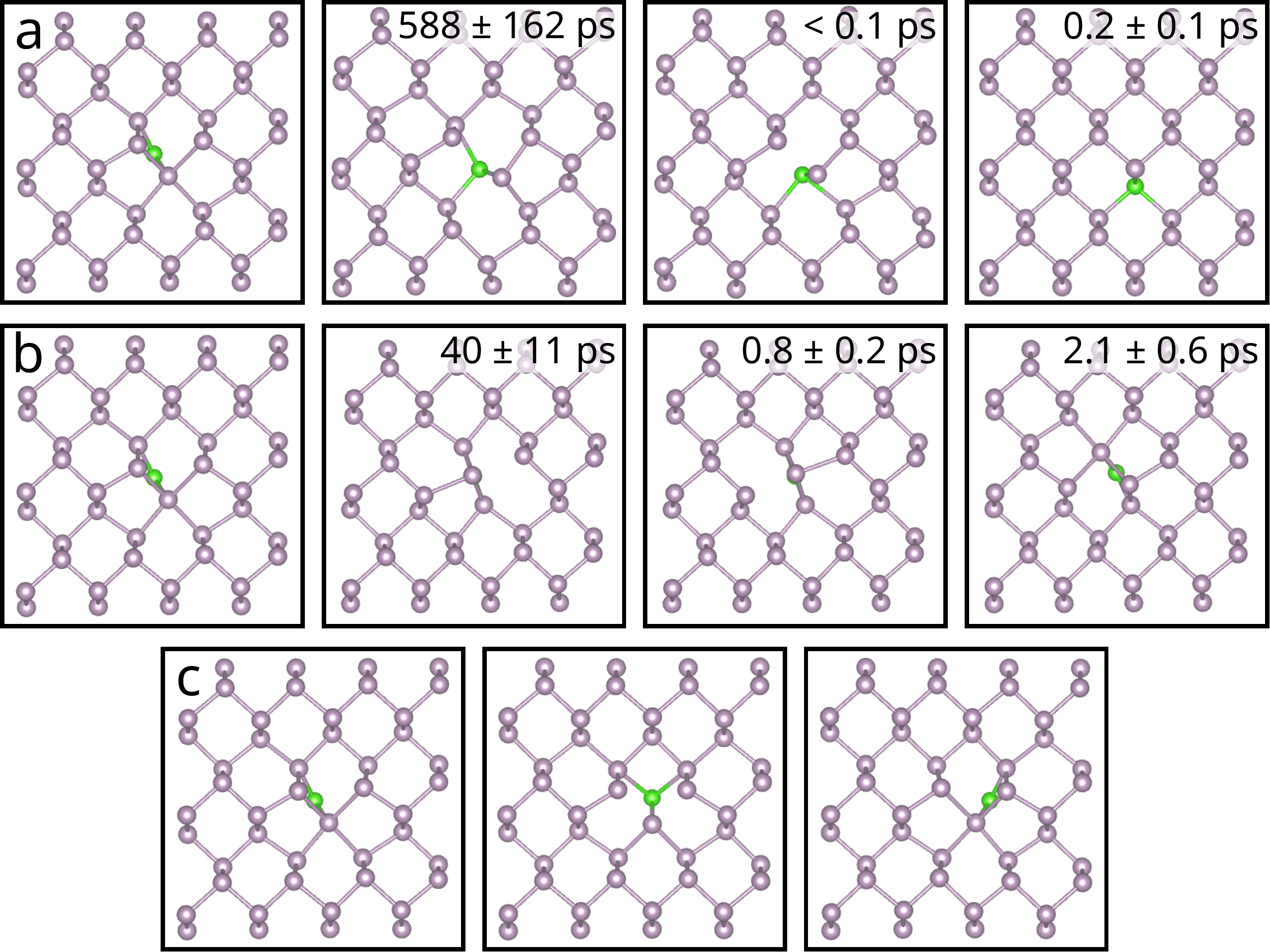}
         \caption{{\bf Transition mechanisms in monolayer phosphorene adatom-vacancy-complexes  observed via HDNNP-MD and ci-NEB.} 
         A recombination process of a monovacancy and an adatom is shown in panel a and two different types of hopping are depicted in panels b and c.
         The hopping in b shows an adatom-vacancy-complex hopping as a whole, while c shows how an adatom changes its orientation with respect to the vacancy.
         All of these transitions start from the same adatom-vacancy-complex configuration.
         The model for this configuration is based on the experimental observation shown in Fig.~\ref{fig:adatoms}g and was used to simulate the corresponding STEM image in Fig.~\ref{fig:adatoms}h.
         Times in the top right corner of each frame in panels a and b indicate the time between the given and the previous frame as observed in HDNNP-MD ($0$~\% strain, $300$~K).
         Light purple atoms indicate phosphorus atoms in the phosphorene structure, while green atoms indicate the phosphorus adatom.
         HDNNP-MD was used to find the transition mechanisms shown in a and b, while ci-NEB calculations were used to identify the transition in c.}
    \label{fig:mechanism}
    \end{figure*}
    
	As for the adatom-vacancy-complexes, we not only observe their creation, but also their recombination, as can be seen in the image series shown in Fig.~\ref{fig:healing}a,b.
	Here, an adatom-vacancy-complex is present in the first frame, which recombines before the second frame is recorded, followed by the creation of two adjacent adatom-vacancy-complexes in the fourth frame which are stable for a few seconds, before recombining to show again a pristine layer in frame six.
	In panels c and d, the number of adatoms and vacancies as a function of electron dose in three distinct image series are shown.
	A clear trend towards the simultaneous creation and annihilation of adatoms and vacancies can be observed, although, some additional vacancies and adatoms are also occasionally seen. 	 
	The observation of more adatoms compared to the number of vacancies might be explained via adatom diffusion from outside the field of view of the analysed images while adatoms adjacent to vacancies almost always stay close to the defect.
	We observe lifetimes of these complexes under electron irradiation between $2.2$~s and $68$~s with a mean lifetime of $19.9 \pm 0.7$~s.
	As the distribution of these lifetimes appears Poissonian, we use the same technique to determine the mean lifetime as for the number of electrons needed to create a single defect.
	Note that lifetimes below $2.2$~s could not be measured as this is the time used to record a single frame.

\begin{figure*}[ht!]
\centering
\includegraphics[width=0.6\textwidth]{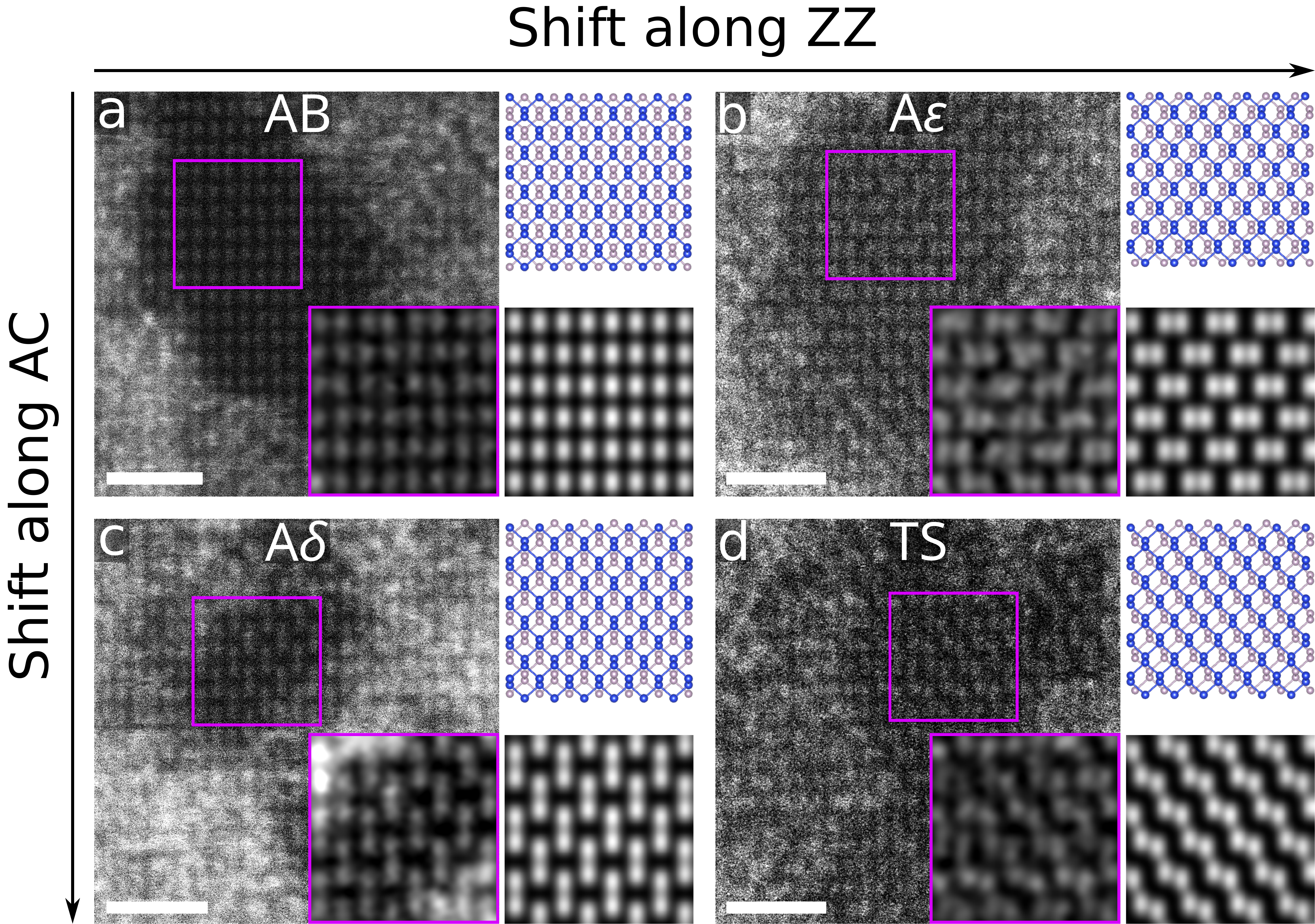}
    \caption{{\bf STEM-HAADF images of different BP stacking orders.} 
    Each panel highlights a different stacking order in bilayer phosphorene.
    The as-recorded image is visible on the left of each panel with a purple square indicating the zoomed-in area shown in the inset. 
    A Gaussian blur was applied to the zoom-in.
    The top right of each panel features a structural model of the different stacking orders from the top view and a STEM-HAADF simulation of this structure is shown on the bottom right.
    Atoms are colored in light purple and blue to indicate the two layers in the bilayer structure.
    Panels a and c were recorded with $1024~\mathrm{px}\times 1024~\mathrm{px}$ and a Gaussian blur of $6$~px was applied in the inset while panels b and d, were recorded with $512~\mathrm{px}\times 512~\mathrm{px}$ and a Gaussian blur of $4$~px was applied in the inset.
    The scale bars are $1$~nm.}
\label{fig:stacking}
\end{figure*}	
    
    We conducted HDNNP-MD simulations under various conditions to investigate the stability of the adatom-vacancy-complex in monolayer phosphorene. 
    Starting from a relaxed structure, similar to the experimental one shown in Fig. \ref{fig:adatoms}g, we tested different conditions, including equilibrium and applied isotropic and anisotropic strains/stresses of 1\% and 3\%. 
    In all cases, we consistently observed recombination, as shown in Fig. \ref{fig:mechanism}a, occurring on average within a timescale of less than one nanosecond.
    Additionally, our simulations revealed two types of hopping mechanisms, one where the entire adatom-vacancy-complex is hopping and one where just the adatom is changing its orientation with respect to the vacancy, as illustrated in Figs.~\ref{fig:mechanism}b and c, respectively. 
    Note that the hopping mechanism shown in Fig.~\ref{fig:mechanism}b was already proposed by theory as a rotation along the ZZ direction for monovacancies in monolayer phosphorene, although no adatom was present for these simulations~\cite{Kyvala2023}.
    Thus, the adatom-vacancy-complex seems to be comparable to a vacancy without an additional adatom in terms of its migration.
    The transition shown in Fig.~\ref{fig:mechanism}c is based on climbing image-nudged elastic band  (ci-NEB) calculations only, revealing an energy barrier of $0.5$~eV for this transition, which could not be verified by HDNNP-MD, as the adatom would always recombine with the vacancy via the transition shown in Fig.~\ref{fig:mechanism}a before the adatom hopping could be observed.
    The validity of these mechanisms was confirmed through DFT calculations. 
    Therefore, the observed stability of the complexes may be due to effects not accounted for in our simulation, such as out-of-equilibrium responses that can not be addressed at the DFT level.
    Possible explanations include a slight bending of the monolayer, an interaction with the bilayer interface, or a charge potential induced by STEM imaging.

\begin{figure*}[ht!]
\centering
\includegraphics[width=0.6\textwidth]{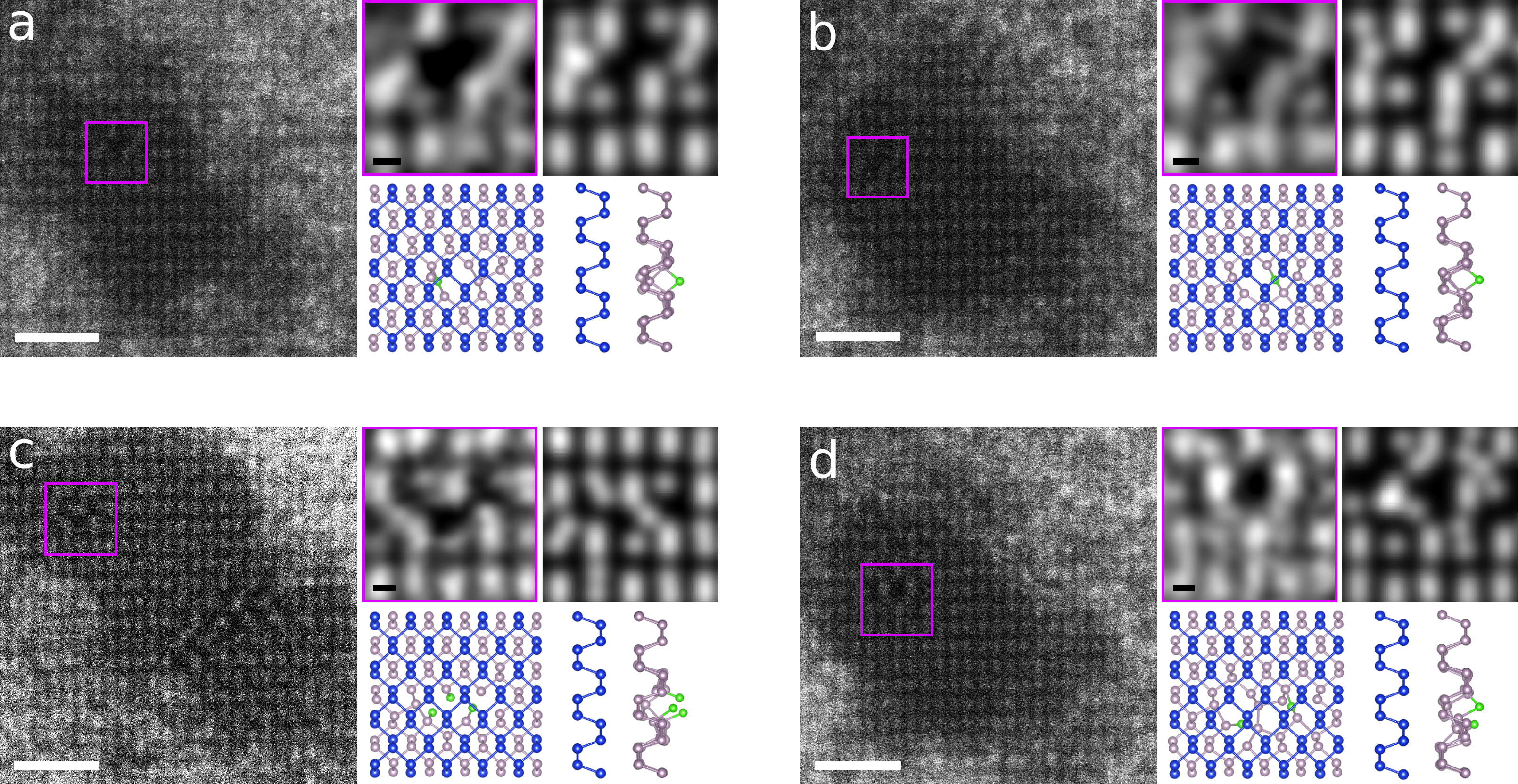}
    \caption{{\bf STEM-HAADF images of different adatom-vacancy configurations along the ZZ direction.} 
    Each panel shows a different defect configuration, including homoatomic adatoms in bilayer phosphorene.
    The as-recorded image is visible on the left of each panel with a purple square indicating the area hosting a defect. 
    A zoom-in of this area is shown after applying a Gaussian blur in the top right, framed with the same color as in the as-recorded image.
    The bottom right of each panel features a structural model of the defect configuration in both top and side view, relaxed using DFT.
    The top right image without a frame is a STEM-HAADF simulation of this model.
    Atoms colored in light purple and blue indicate phosphorus atoms in the two different layers of the bilayer structure and green atoms indicate phosphorus adatoms, respectively.
    Panels a, b, d were recorded with $512~\mathrm{px}\times 512~\mathrm{px}$ and a Gaussian blur of $4$~px was applied while panel c, was recorded with $1024~\mathrm{px}\times 1024~\mathrm{px}$ and a Gaussian blur of $6$~px was applied.
    The scale bars are $1$~nm in the as-recorded images and $1$~\r{A} in the zoomed-in images.}
\label{fig:configurationsZZ}
\end{figure*}	    

    During measurements where the phosphorene sample was continuously irradiated with highly energetic electrons, we could also observe local shifts in the stacking order of the bilayer.
    Four distinct stacking orders could be identified, as shown in Fig.~\ref{fig:stacking}.
    Panel a shows the known AB stacking order that was reported both for bulk BP~\cite{hultgren_atomic_1935} as well as thin layers~\cite{wu_identifying_2015}.
    The other stacking orders are characterized by a shift along the ZZ direction for the structure labeled here A$\varepsilon$ (Fig.~\ref{fig:stacking}b), a shift along the AC direction for the A$\delta$ structure (Fig.~\ref{fig:stacking}c), and a shift along both directions for the TS structure (Fig.~\ref{fig:stacking}d).
    The A$\delta$ and TS stacking order have also been studied theoretically, showing their (meta-)stability as well as their influence on the electronic structure of the phosphorene bilayer~\cite{alhassan_insight_2024}, while we could not find any reports on the A$\varepsilon$ structure in the literature.
    One noteworthy theoretical prediction for a transition from the AB into the TS or A$\delta$ stacking order is the induced transition from a direct to an indirect band gap, which is otherwise absent in BP. 
    These findings mark the first experimental observation of these different stacking orders in bilayer phosphorene. 

\begin{figure*}[ht!]
\centering
\includegraphics[width=0.6\textwidth]{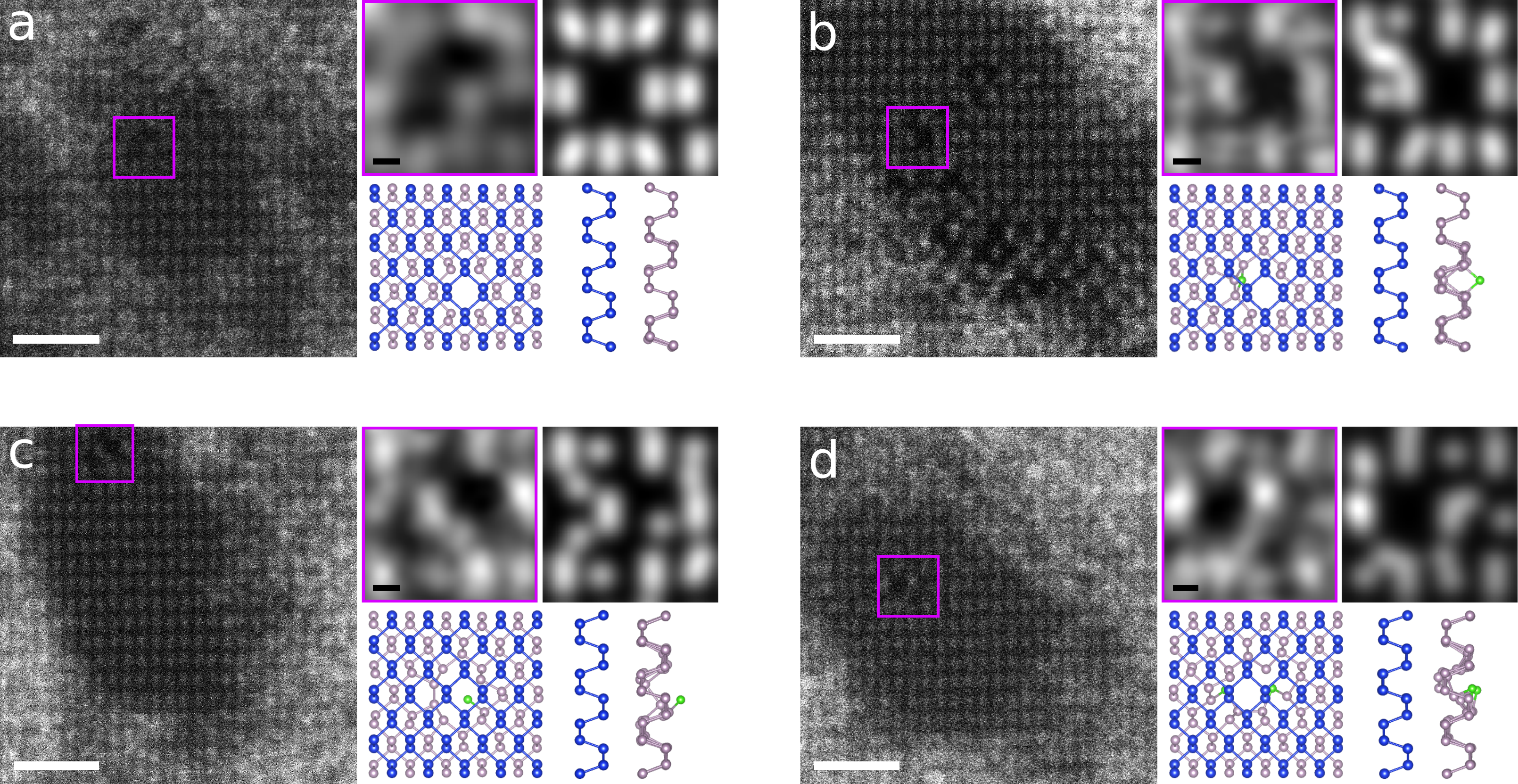}
    \caption{{\bf STEM-HAADF images of different adatom-vacancy configurations along the AC direction.} 
    Each panel shows a different defect configuration, including homoatomic adatoms in bilayer phosphorene.
    The as-recorded image is visible on the left of each panel with a purple square indicating the area hosting a defect. 
    A zoom-in of this area is shown after applying a Gaussian blur in the top right, framed with the same color as in the as-recorded image.
    The bottom right of each panel features a structural model of the defect configuration in both top and side view, relaxed using DFT.
    The top right image without a frame is a STEM-HAADF simulation of this model.
    Atoms colored in light purple and blue indicate phosphorus atoms in the two different layers of the bilayer structure and green atoms indicate phosphorus adatoms, respectively.
    Panels a, b, c were recorded with $1024~\mathrm{px}\times 1024~\mathrm{px}$ and a Gaussian blur of $6$~px was applied while panel d, was recorded with $512~\mathrm{px}\times 512~\mathrm{px}$ and a Gaussian blur of $4$~px was applied.
    The scale bars are $1$~nm in the as-recorded images and $1$~\r{A} in the zoomed-in images.}
\label{fig:configurationsAC}
\end{figure*}	

	Finally, we want to further highlight the variety of different defect configurations beyond monovacacies (see Fig.~\ref{fig:adatoms}) observed in phosphorene bilayers. 
    Fig.~\ref{fig:configurationsZZ} shows a divacancy (panel a), trivacancy (panel b) and two different configurations of tetravacancies (panels c and d) along the ZZ direction with a variable number of adatoms. 
    Fig.~\ref{fig:configurationsAC} shows the same sequence of defects (divacancy in panel a, trivacancy in panel b and two configurations of tetravacancies in panels c and d), this time orientated along the AC direction.
	Note that the structures in Fig.~\ref{fig:configurationsZZ}a and Fig.~\ref{fig:configurationsZZ}d have been previously reported without adatoms~\cite{yao_situ_2020, rabiei_baboukani_defects_2022}, while the other structures are yet unreported.
    With the exception of Fig.~\ref{fig:configurationsAC}a, a reasonably good agreement between the simulated and the experimentally observed structures could be achieved, indicating the correct identification of these defects.
    The discrepancies for Fig.~\ref{fig:configurationsAC}a might be explained by the occurrence of A$\delta$ stacking (see Fig.~\ref{fig:stacking}c) in the experimental image, which might influence the stability of the defect structure.

\section*{Conclusions}

	In conclusion, we characterized a number of yet unreported defect configurations in mono- and bilayer phosphorene, including stable adatom-vacancy-complexes.
	The experimentally obtained displacement cross section of bilayer phosphorene under $60$~kV electron irradiation was measured to be $7.7 \pm 1.4$~barn with most defects being located in just one single layer.
	Within this layer, the two atomic planes also showed clearly distinct cross section values ($10.4 \pm 1.6$~barn and $2.0 \pm 0.3$~barn, presumably for the planes away and towards the electron source, respectively).
	The average lifetime of adatom-vacancy-complexes was found to be $19.9 \pm 0.7$~s with a maximum observed lifetime of up to $68$~s.
    Using DFT, the observation of different adatom-vacancy configurations was verified while the extended lifetime of the adatom-vacancy-complexes could not be reproduced using HDNNP-MD, leaving the explanation for the stability of these structures as an open question that requires further investigation.
    These results further extend the understanding of the creation and stability of defects in phosphorene, which might lead to the creation of tailored defect-engineered phosphorene devices.

\section*{Methods}

    {\bf Sample preparation} Phosphorene samples were produced inside an Ar glovebox with both O$_2$ and H$_2$O partial pressures below $0.5$~ppm using the mechanical exfoliation technique introduced by Novoselov et al.~\cite{novoselov_electric_2004}. 
    The bulk BP crystal was commercially bought from HQ Graphene.
    After exfoliation, thin flakes of BP were transferred onto a custom-made $200$~$\mu$m thick silicon TEM grid with 3$\times$3 windows covered by a $1$~$\mu$m thick holey silicon nitride membrane using an all-dry visocoelastic stamping method~\cite{castellanos-gomez_deterministic_2014}.
    Each window has a 5$\times$5 matrix of 2~$\mu$m holes with a separation of 5~$\mu$m between holes.
    Samples were introduced to ultra-high vacuum (UHV) within $20$~min after the transfer to the TEM grids was finished, and they were baked over night at ca. 150$^\circ$C before any measurements were conducted.
    Due to the unique design of the CANVAS ultra-high vacuum (UHV) system~\cite{mangler_materials_2022}, connecting the glovebox directly to the loadlock of the system, the samples could be transferred into vacuum (pressure $\sim$~10$^{-9}$~mbar) without exposure to ambient conditions, avoiding oxidation of the material to influence the experimental results.
    In between measurements, samples were kept in vacuum to prevent degradation.

    {\bf Scanning transmission electron microscopy} The microscope used for this study is a Nion UltraSTEM 100 equipped with a cold field emission gun obtaining a probe size of $\sim 1$~\r{A}.
     Images were recorded at room temperature using a HAADF detector with collection angles of $80-300$~mrad at a constant electron acceleration voltage of $60$~kV and a beam convergence semiangle of $30$~mrad.
     Additionally, the number of impinging electrons on the sample was measured as explained in Ref.~\cite{speckmann_combined_2023} resulting in a beam current of $16.5 \pm 0.5$~pA.
     The base pressure the samples were exposed to inside the microscope column was below 10$^{-9}$~mbar at all times. 
     
    {\bf Computational methods} First-principles calculations were performed using the Vienna Ab Initio Package~\cite{vasp1,vasp2}, employing the same setup as in our previous study~\cite{Kyvala2023} with optimised lattice parameters for pristine phosphorene samples of $3.30$~\r{A} (along ZZ) and $4.62$~\r{A} (along AC) for the monolayer and $3.30$~\r{A} (along ZZ) and $4.50$~\r{A} (along AC) for the bilayer, respectively.
    For dynamic studies, we utilize the high dimensional neural network potential (HDNNP)~\cite{Behler2007}. 
    This potential has been successfully employed for simulating the diffusion of monovacancies and coalescence of monovacancies in phosphorene in our previous work~\cite{Kyvala2023}. 
    Given that our current study deals with a similar problem, we employ the same set of descriptors, consisting of 50 symmetry functions, which characterize the local chemical environment of phosphorene up to a cutoff distance of $7$~\r{A}. 
    The neural network architecture consists of two hidden layers, each with 40 nodes, utilizing the softplus activation function. 

    To discover the recombination mechanism, we employ an active learning strategy called query by committee~\cite{Seung1992} and iteratively improve our neural network potential. 
    This strategy has been proven effective in creating training sets for machine learning potentials~\cite{Smith2018,Schran2020,Loeffler2020,Smith2021}, including those for phosphorene~\cite{Kyvala2023}. 
    The approach involves training several HDNNPs with different training/validation splits and initial conditions. 
    During molecular dynamics (MD) simulation, we monitor the energy and force predictions of these committee members to identify configurations where predictions disagree. 
    When the committee disagreement exceeds a predetermined threshold, we stop MD simulation accelerated by HDNNPs and continue it with {\it ab initio} MD, generating new configurations that are incorporated into the training set. 
    The neural networks are then retrained on this enlarged dataset, and the MD simulation is resumed until the desired level of accuracy is achieved. 
    The committee of 8 HDNNPs reproduces the energies and forces of the training set with a root mean squared error (RMSE) of $0.7$~meV/atom and $51.3$~meV/\r{A}, respectively. 
    The reference data, comprising 4904 configurations, and the trained models are publicly available on Zenodo~\cite{zenodo}. 
    
    The HDNNPs used in this work were constructed and trained with the n2p2 package~\cite{Singraber2019a}. 
    All HDNNP-MD simulations were conducted using the Large-scale Atomic/Molecular Massively Parallel Simulator (LAMMPS)~\cite{Thompson2022}. 
    We employed the microcanonical ensemble (NVE) and the velocity Verlet integrator with a time step of $1$~fs.

    STEM image simulations were performed using {\it ab}TEM~\cite{madsen_abtem_2021} and structure models shown in the figures were created with VESTA~\cite{momma_vesta_2011}.

	{\bf Data analysis} Data was analysed manually by identifying the different vacancy and adatom sites within each frame of each image series individually.
	Similar to Ref.~\cite{speckmann_combined_2023}, the uncertainty for this process will result in an increased uncertainty of the cross sections calculated with those data.
	This is taken into account by increasing the uncertainty obtained by fitting the distribution of defects to Poisson statistics by a factor of $2.5$.
	Other uncertainties were deemed insignificant compared to this.
	To not skew the distribution of created defects by the simultaneously occurring recombination processes, image series were cut off when recombination was observed, while diffusion of defects in between frames was assumed when the number of defects in subsequent frames was identical, but not their positions.
	For the estimation of the sublayer cross sections, defects in each image series were sorted according to their occurrence in the two different planes.

\section*{Acknowledgments}

We acknowledge funding from the Austrian Science Fund (FWF) through the doctoral college Advanced Functional Materials--Hierarchical Design of Hybrid Systems (DOC~85), and via the projects 10.55776/DOC142 and 10.55776/P35318 as well as from the Vienna Doctoral School in Physics.
The computational results presented have been achieved in part using the Vienna Scientific Cluster (VSC).

\bibliography{references}

\end{document}